\documentclass{llncs}

\usepackage{mathpazo}
\usepackage{microtype}
\usepackage{graphicx} %
\usepackage{tikz}
\usepackage{amssymb}
\usepackage{amsmath}
\usepackage{hyperref}
\hypersetup{pdfauthor = {Neil Dalchau, Niall Murphy, Rasmus Petersen, Boyan Yordanov},
		    pdftitle = {Synthesizing and tuning chemical reaction networks with specified behaviours},
		    colorlinks = true,
		    urlcolor = black,
		    linkcolor = black,
		    citecolor = black}

\newcommand{\Integers}{\mathbb{Z}}
\newcommand{\Naturals}{\mathbb{N}_0}
\newcommand{\Reals}{\mathbb{R}}
\newcommand{\Bool}{\mathbb{B}}
\newcommand{\figref}[1]{{Fig.~\ref{#1}}}
\usepackage{verbatim}

\usepackage{complexity}

\usepackage{enumitem}
\newlist{compactitem}{itemize}{3}
\setlist[compactitem]{topsep=0pt,partopsep=0pt,itemsep=0pt,parsep=0pt}
\setlist[compactitem,1]{label=\textbullet}
\setlist[compactitem,2]{label=---}
\setlist[compactitem,3]{label=*}

\newlist{compactenum}{enumerate}{3}
\setlist[compactenum]{topsep=0pt,partopsep=0pt,itemsep=0pt,parsep=0pt}
\setlist[compactenum,1]{label=\arabic*}
\setlist[compactenum,2]{label=\alph*}
\setlist[compactenum,3]{label=\roman*}

\newcommand{\prd}{\mathbf{p}}
\newcommand{\reac}{\mathbf{r}}
\newcommand{\rates}{\mathbf{k}}
\newcommand{\rate}{k}

\DeclareRobustCommand\filledcirc{\tikz \fill[black] (0,0) circle (.5ex); }

\title{Synthesizing and tuning chemical reaction networks with specified behaviours}
\titlerunning{Synthesizing CRNs}
\author{Neil Dalchau, Niall Murphy, Rasmus Petersen, Boyan Yordanov}
\authorrunning{Dalchau et al.,}
\institute{Microsoft Research, Cambridge, CB1 2FB, UK,
\email{\{ndalchau,a-nimurp,a-rapete,yordanov\}@microsoft.com}}

\begin{document}

\maketitle

\begin{abstract} We consider how to generate chemical reaction networks (CRNs) from functional specifications. 
We propose a two-stage approach that combines synthesis by satisfiability modulo theories and Markov chain Monte Carlo based optimisation. 
First, we identify candidate CRNs that have the \emph{possibility} to produce correct computations for a given finite set of inputs. %
We then optimise the reaction rates of each CRN using a combination of stochastic search techniques applied to the chemical master equation, simultaneously improving the \emph{probability} of correct behaviour and ruling out spurious solutions.
In addition, we use techniques from continuous time Markov chain theory to study the expected termination time for each CRN. 
We illustrate our approach by identifying CRNs for majority decision-making and division computation, which includes the identification of both known and unknown networks.

\end{abstract}

\section{Introduction}
A central goal of molecular programming is to be able to implement arbitrary dynamical behaviours.
Chemical reaction networks (CRNs) are a popular formalism for describing
biochemical systems, such as protein interaction networks, gene regulatory
networks, synthetic logic circuits and molecular programs built from DNA.
Extensive theoretical understanding exists about the behaviour of a multitude
of CRNs, and the behaviour of some networks has been exhaustively explored~\cite{Wilhelm2009}.
Besides describing chemical systems, CRNs provide a common language for expressing problems studied in computer science theory (e.g. Petri nets, network protocols) as well as control theory and engineering. Methods exist to convert CRNs into equivalent physical implementations, based on DNA strand displacement~\cite{SSW2010,Chen2013} the DNA toolbox system~\cite{Fujii2013} and genelets~\cite{Kim2011}.
Therefore, we sought to develop a methodology for proposing candidate CRNs that exhibit a pre-specified behaviour.

The computational power of CRNs has been extensively studied~\cite{Cook2009}.
It is known that error-free (stably computing~\cite{AADFP2006}) CRNs compute
exactly the class of semi-linear functions~\cite{CDS2012,AAE2006b}. %
However, if the stability restriction is relaxed and we allow the CRN to
sometimes compute the wrong answer then it is possible to implement a register
machine, that is, CRNs with error can compute functions beyond the
semi-linear class (indeed they are equivalent in power to Turing
machines)~\cite{AAE2006,Cook2009}.

Although there are procedures to generate CRNs for semi-linear
functions \cite{CDS2012,AAE2006}, primitive recursive functions~\cite{Cook2009},
or even from arbitrary Turing machines~\cite{Cook2009},
the proposal of practical (i.e.\ experimentally implementable) CRNs that
compute a given function has thus far mostly been a manual effort.
In this work, we attempt to automate the proposal of CRNs, by formally specifying a behaviour and automatically identifying CRNs that satisfy the desired behaviour with high probability.
First, we formalise the problem of identifying CRNs that have the capacity to
produce correct, finite computations for a given finite set of inputs. This
corresponds to a synthesis problem, as opposed to verification, where the goal
is to determine the correctness of a given CRN~\cite{Yordanov2013}. We express
CRN synthesis as a satisfiability modulo theories (SMT) problem, which can be
addressed using solvers such as Z3~\cite{deMoura2008}. This allows us to
generate a number of canditate CRNs or to prove that no such CRN of a given size
(in terms of numbers of reactions, species and computation lengths) exists.
However, while the existence of correct computations is guaranteed for each generated CRN, the probability of these computations might be low.

To determine whether correct computations can occur with high probability, we next optimise the reaction rates of each generated CRN.
To solve the optimisation problem, we combine stochastic search strategies based on Markov chain Monte Carlo (MCMC) with numerical integration of the chemical master equation (CME). This part of the problem was recently addressed in~\cite{Han2008,Ceska2014}, though applied only to a single input.

In this paper, we specifically focus on uniform CRNs, those that have a fixed number of species and reactions for all input sizes.
We also restrict our attention to bimolecular CRNs, where there are precisely 2 reactants and 2 products in every reaction.
Bimolecular CRNs are equivalent to Population Protocols (PPs)~\cite{AADFP2006} and also guarantee that mass is conserved in the system.
We applied our two-step approach first to majority decision-making, in which the network seeks to identify which of two inputs is in an initial majority.
Majority networks are well-studied in the literature, and there are many known CRNs that give approximate solutions~\cite{Angluin2008,PVV2009,Cardelli2014}.
We then applied our approach to division, a non-linear function which has been relatively less studied.
We show a range of CRNs for majority and division identified automatically using our method, some of which have been identified and characterised previously, though some of which are entirely novel. This illustrates the potential for automatically determining CRNs with a specified behaviour.

\section{Preliminaries}\label{sec:background}

A chemical reaction network (CRN) is a tuple $\mathcal{C} = (\Lambda,\mathcal{R})$, where $\Lambda = \{s_0, \ldots, s_n\}$ and $\mathcal{R} = \{r_0, \ldots, r_m\}$ denote the finite sets of species and reactions, respectively. A reaction is a tuple $r = (\reac^r, \prd^r, \rate^r)$ where $\reac^r$ and $\prd^r$ are the reactant and product {\em stoichiometry} vectors ($\reac^r_s \in \Naturals$ and $\prd^r_s \in \Naturals$ denote the stoichiometry of each species $s \in \Lambda$), $\rate^r \in \Reals_{\geq0}$ denotes the rate of $r$ and $\rates$ denotes the vector of all reaction rates. Given a reaction $r = (\reac^r, \prd^r, \rate^r)$, the set of reactants of $r$ is $\{s \in \Lambda\;|\;\reac^r_s>0\}$ and the set of products of $r$ is $\{s \in \Lambda\;|\;\prd^r_s>0\}$.
%
In this paper, we focus on the class of {\em bimolecular} CRNs, where $\sum_{s \in \Lambda} \reac^r_s = 2$ and $\sum_{s \in \Lambda} \prd^r_s = 2$, for all reactions $r \in \mathcal{R}$.%

The dynamical behaviour of bimolecular CRNs can be understood as follows.
The set of all possible system states is $X = \Naturals^{|\Lambda|}$, where a state $x \in \Naturals^{|\Lambda|}$ represents the number of molecules of each species.
We denote the number of molecules of species $s \in \Lambda$ at state $x$ by $x_s$.
Given a reaction $r\in \mathcal{R}$ where $\reac^r_s=2$ for some $s \in \Lambda$, the {\em propensity}%
\footnote{We assume that the reaction volume is 1 to allow for later volume scaling
  e.g.\ $\rate_x^r/v$ is the propensity for a reaction volume equal to $v$}
of $r$ at $x$ is $\rate_x^r = \rate^r\cdot \frac{x_s \cdot (x_s-1)}{2}$.
If, on the other hand, $\reac^r_s=\reac^r_{s'}=1$ for some species $s,s'$, the propensity of $r$ is $\rate_x^r = \rate^r\cdot x_s \cdot x_{s'}$.
The time at which reaction $r$ would fire, once the system enters state $x \in X$, is stochastic and follows an exponential distribution with a rate determined by the reaction's propensity $\rate_x^r$.
Assuming that reaction $r$ is the first one to fire, the state of the system is updated as $x'_s = x_s - \reac^r_s + \prd^r_s$ for all $s \in \Lambda$, where $x$ and $x'$ are the current and next states.

An abstraction of CRNs that preserves reachability but does not consider reaction rates or time is given by the {\em transition system} $\mathcal{T}^\mathcal{C} = (X, T)$, where the transition relation $T$ is defined as
\begin{equation}\label{eqn:trans}
\forall x,x' \in X\;.\; T(x,x') \leftrightarrow \bigvee_{r \in \mathcal{R}} \bigwedge_{s \in \Lambda} \left( x_s \geq \reac^r_s \wedge x'_s = x_s - \reac^r_s + \prd^r_s\right).
\end{equation}
In other words, the choice between reactions from $\mathcal{R}$ is non-deterministic but enough molecules of each reactant must be present in state $x$ for the reaction to fire.
The transition between states $x$ and $x'$ happens when any reaction $r\in \mathcal{R}$ fires and the number of molecules is updated accordingly. A path $x_0,x_1,\ldots$ of $\mathcal{T}$ satisfies $T(x_i,x_{i+1})$ for $i=0,1,\ldots$ and, given an initial state $x_0$ we call state $x_f$ reachable from $x_0$ if there exists a path $x_0, \ldots, x_f$.

Given a CRN $\mathcal{C}$, let $X_0 \subseteq X$ denote a finite set of initial states and $X_r \subseteq X$ denote the set of states reachable from $X_0$. Assuming that $X_r$ is finite, $\mathcal{C}$ can be represented as a \emph{continuous time Markov chain} (CTMC) that preserves information about the transition probabilities and rates that determine the stochastic behaviour of the system and the expected execution times.
We define a CTMC to be a tuple $\mathcal{M} = (X_r, \pi_0, \vec{Q})$, where $X_r$ is a finite set of states, $\pi_0:X_r\rightarrow\mathbb{R}$ is the initial distribution of molecule copy numbers of all species, and $\vec{Q}:X_r\times X_r \rightarrow\mathbb{R}$ is a matrix of transition propensities. While the set of initial states is not represented explicitly, it is captured through the initial distribution, i.e. $X_0 = \{x \in X_r\;|\; \pi_0(x)>0\}$. A CTMC $\mathcal{M}^\mathcal{C}$ is constructed from a CRN $\mathcal{C}$ by first determining the set of reachable states, and then evaluating the propensities of each reaction. The $(i,j)^\text{th}$ entry of $\vec{Q}$, $q_{ij}$, represents a transition from state $x_i$ to state $x_j$. Accordingly, $q_{ii}$ is the remaining probability mass, equal to $-\sum_{i\neq j}q_{ij}$. The transient probability vector $\pi_t$ evolves according to $\frac{d\pi_t}{dt} = \pi_t\vec{Q}$, which is known as the chemical master equation (CME).

Following \cite{Han2008,Ceska2014}, a parametric CTMC (pCTMC) is a CTMC where the reaction rates are parameterised by $\rates$, as above. Denote by $\mathcal{P}$ the parameter space, $\mathcal{P}:\Reals_{\geq0}^{P}$, such that $\rates$ is instantiated by a parameter point $p\in\mathcal{P}$. Accordingly, given a pCTMC $\mathcal{M}$ and parameter space $\mathcal{P}$, an instantiated pCTMC $\mathcal{M}_p = (X,\pi_0,\vec{Q}_p)$ is an evaluation at point $p\in\mathcal{P}$.

\section{Problem formulation}
\label{sec:problem}

The main problem we consider in this paper, which we formalise in this section, is the identification of CRNs that satisfy given properties. Specifically, we are interested in finite reachability properties, which capture a range of interesting CRN behaviours.

Let $\mathcal{C} = (\Lambda,\mathcal{R})$ be a given CRN and $\mathcal{T}^\mathcal{C} = (X, T)$ and $\mathcal{M}^\mathcal{C} = (X_r, \pi_0, \vec{Q})$ denote its transition system abstraction and CTMC representation, as discussed in Section~\ref{sec:background}. Let $\phi : X \to \Bool$ denote a {\em state predicate}, constructed using
\begin{eqnarray*}
\phi& ::=   &   E_b   \\
E_b & ::=   &   \textit{true}\;|\;\textit{false}\;|\; E_c\;|\;\neg E_b\;|\;E_b \rhd E_b \mbox{ where } \rhd \in \{\wedge, \vee, \Rightarrow, \Leftrightarrow\} \\
E_c & ::=   &   E_a \rhd E_a \mbox{ where } \rhd \in \{<,\leq, =, >, \geq \} \\
E_a & ::=   &   s \in \Lambda\;|\;c \in \Integers\;|\;E_a \rhd E_a \mbox{ where } \rhd \in \{+,-,*\}.
\end{eqnarray*}
For example, if $\phi := s>5$, then $\phi(x)$ denotes that $x_s>5$.

In this paper, we consider {\em path predicates} $\Phi = (\phi_0,\phi_F)$, which are expressed using two state predicates that must be satisfied at the initial ($\phi_0$) and at some final ($\phi_F$) state of a path. Let $K$ denote the number of steps we consider.
\begin{definition}\label{def:ts_sat}
Given a finite path $\rho: x_0\ldots x_K$ of $\mathcal{T}^\mathcal{C}$ we say
that $\rho$ satisfies path predicate $\Phi = (\phi_0,\phi_F)$, denoted as
$\rho \vDash \Phi$, if and only if $\phi_0(x_0) \wedge \phi_F(x_K)$ evaluates
to true and no reactions are enabled in $x_K$ (i.e. $x_K$ is a terminal
state)%
\footnote{We consider terminating computations by enforcing that no reactions
  are enabled at the state that satisfies $\phi_F$.
  Alternative strategies possible within our approach could consider reaching
  a fix-point (i.e. the firing of any enabled reaction does not cause a
  transition to a different state), or reaching a cycle along which $\phi_F$
  is satisfied, to guarantee that the correct output is eventually computed
  and remains unchanged by any subsequent reactions.}.
\end{definition}

We define the probability of $\Phi$, denoted $P_\Phi$, using $\mathcal{M}^\mathcal{C}$ as follows. Let $X_0 = \{x \in X\;|\;\phi_0(x)\}$ denote the set of states that satisfy the initial state predicate. We initialise $\mathcal{M}^\mathcal{C}$ with a uniform sample from the states that satisfy $\phi_0$, which defines $\pi_0$ as
\begin{equation*}
\pi_0(x) =
\left\{
\begin{array}{l l}\frac{1}{|X_0|} & \mbox{ if } x \in X_0\\
 0 & \mbox{ otherwise}
\end{array}
\right.
\end{equation*}
Similarly, $X_F = \{x \in X\;|\;\phi_F(x)\}$ denotes the set of states satisfying the final state predicate.
\begin{definition}\label{def:prob}
The probability of $\Phi$ is defined as
\begin{equation*}
P_\Phi = \sum_{x \in X_F} \pi_t(x),
\end{equation*}
where $t$ denotes the maximal time we consider and $\pi_t$ is the probability
vector at time $t$ computed using the CME introduced in
Section~\ref{sec:background}. In other words, we define $P_\Phi$ as the
average probability of the states satisfying $\phi_F$ at time $t$.
\end{definition}
Note that it is possible to optimise for both speed and accuracy by, for
example, defining $P_\Phi$ to be the integration of the
probability mass of all states satisfying $\phi_F$ from time 0 to time $t$.

\begin{problem}\label{prob:main}
Given a finite set of path predicates $\{\Phi_0,\ldots, \Phi_n\}$, find a bimolecular CRN $\mathcal{C}$ such that
\begin{enumerate}
\item \label{prob:first} for each $\Phi_i$, there exists a path $\rho_i$ of $\mathcal{T}^\mathcal{C}$, such that $\rho_i \vDash \Phi_i$ and
\item \label{prob:secnd} the average probability $\frac{\sum_{i=0}^{n} P_{\Phi_i}}{n+1}$ defined using $\mathcal{M}^\mathcal{C}$ is maximised.
\end{enumerate}
\end{problem}

\section{Synthesis and tuning of CRNs}

We solve Problem~\ref{prob:main} by addressing each of the two subproblems
separately. First, we generate a number of CRNs that satisfy the
specifications from Problem~\ref{prob:main}.\ref{prob:first} using a
satisfiability modulo theories (SMT)-based approach (Section~\ref{sec:smt}). The CRNs identified at that point are capable of
producing a path that satisfy each path predicate, which addresses
Problem~\ref{prob:main}.\ref{prob:first} but they might also include incorrect
paths and the probability of correct computations might be low. Therefore, we
tune the reaction rates of these CRNs in order to maximise the average
probability (discussed in Section~\ref{sec:optim}), which
addresses Problem~\ref{prob:main}.\ref{prob:secnd}.

\subsection{SMT-based synthesis}\label{sec:smt}
Here, we present our approach to finding a bimolecular CRN $\mathcal{C}$ that satisfies a specification expressed as path predicates $\{\Phi_0,\ldots, \Phi_n\}$ (Problem~\ref{prob:main}.\ref{prob:first}). We address this problem by encoding $\mathcal{T}^\mathcal{C}$ symbolically for any possible bimolecular CRN $\mathcal{C} = (\Lambda,\mathcal{R})$ where $|\mathcal{R}| = M$ and $|\Lambda| = N$ (i.e. the number of species and reactions is given), together with the specification $\{\Phi_0,\ldots, \Phi_n\}$ for some finite number of steps $K$, as a satisfiability modulo theories (SMT) problem. We then use the SMT solver Z3 \cite{deMoura2008} to enumerate bimolecular CRNs that satisfy the specification or prove that no such CRNs exists for the given $N$, $M$, and $K$. Finally, we apply an incremental procedure to search for CRNs of increasing complexity (larger $N$ and $M$) or to provide more complete results by increasing~$K$.

Using Z3's theory of linear integer arithmetic, we represent the stoichiometry of $\mathcal{C}$ as two symbolic matrices $\reac \in \Naturals^{M\times N}$ and $\prd \in \Naturals^{M\times N}$ (using integer constraints to prohibit negative integers). Given a reaction $r \in \mathcal{R}$ and species $s \in \Lambda$, $\reac^r_s$ ($\prd^r_s$) defined in Section~\ref{sec:background} is now encoded as a symbolic integer.
We ensure that only bimolecular CRNs are considered by asserting the constraints $\bigwedge_{i=0}^{M-1}\sum_{j=0}^{N-1}\reac_{i,j} = 2$ and $\bigwedge_{i=0}^{M-1}\sum_{j=0}^{N-1}\prd_{i,j} = 2$. In addition, we introduce the following constraints.
\begin{itemize}
\item We label a subset of the species $\Lambda_I \subseteq \Lambda$ as inputs and assert that $\bigwedge_{s \in \Lambda_I} \bigvee_{r \in \mathcal{R}} \reac^r_s > 0$ to ensure all inputs are consumed by at least one reaction.
\item We label a subset of the species $\Lambda_O \subseteq \Lambda$ as outputs and assert that $\bigwedge_{s \in \Lambda_O} \bigvee_{r \in \mathcal{R}} \prd^r_s > 0$ to ensure all outputs are produced by at least one reaction.
\item We assert that $\bigwedge_{r,r' \in \mathcal{R}, r\neq r'} \bigvee_{s \in \Lambda} \prd^r_s\neq \prd^{r'}_s \vee \reac^r_s\neq \reac^{r'}_s$ to ensure that two reactions never have the same reactants and products and, therefore, all $M$ reactions are utilised.
\item Finally, we assert that $\bigwedge_{r \in \mathcal{R}} \bigvee_{s \in \Lambda} \prd^r_s \neq \reac^r_s$ to ensure that the firing of each reaction updates the state of the system.
\end{itemize}

Following an approach inspired by bounded model checking (BMC)~\cite{Biere1999}, we represent the finite path $\rho_i = x^i_0,\ldots x^i_K$ for each $\Phi_i$ by defining each state as a symbolic vector $x^i_j \in \Naturals^{N}$ and ``unrolling" the transition relation of $\mathcal{T}_\mathcal{C}$ (i.e. asserting the constraint $T(x^i_j,x^i_{j+1})$ for each $i = 0\ldots n$ and $j=0\ldots K-1$). For each path predicate $\Phi_i = (\phi_0, \phi_F)$ and path $\rho_i$ we then assert the constraint $\phi_0(x_0^i) \wedge \phi_F(x_K^i) \wedge \textit{Terminal}(x_K^i)$ according to Def.~\ref{def:ts_sat}, where $\textit{Terminal}(x) \triangleq \bigwedge_{r \in \mathcal{R}} \bigvee_{s \in \Lambda} x_s < \reac^r_s$, i.e. no reactions are possible due to insufficient molecules of at least one reactant.

The parameter $K$ specifies the maximal trajectory length that is considered.
The BMC approach is conservative, since computations that require more than~$K$
steps (reaction firings) to reach a state satisfying $\phi_F$ will not be
identified. Increasing $K$ leads to a more complete search, and indeed the
approach becomes complete for a sufficiently large $K$ determined by the
diameter of a system, but also increases the computational burden.
To alleviate this, we follow an approach from \cite{Yordanov2013}
and consider \emph{stutter} transitions (corresponding to multiple firings of
the same reaction in a single step) by using the following modified transition
relation definition $T_{st}$ (as opposed to $T$ from Eqn.~\ref{eqn:trans})
\begin{align*}\label{eqn:trans_stutter}
&\forall x,x' \in X\;.\; T_{st}(x,x') \leftrightarrow (\textit{Terminal}(x) \wedge x = x')\; \vee \\
&\quad
 \exists n \in \mathbb N\;.\;
 \bigvee_{r \in \mathcal{R}}
  \bigwedge_{s \in \Lambda}
   \left( x_s \geq \reac^r_s \wedge
          x_s \geq n\cdot(\reac^r_s - \prd^r_s) \wedge
          x'_s = x_s + n\cdot(\prd^r_s - \reac^r_s)
   \right).
\end{align*}
For any enabled reaction $r$ ($x_s \geq \reac^r_s$), $T_{st}$ allows $r$ to
fire up to $n$ times in the stutter transition.
$n$ is limited by the consumption and production of the species needed for the
reaction to fire ($x_s \geq n\cdot(\reac^r_s - \prd^r_s)$).
In many cases, stutter transitions dramatically decreases the required trajectory lengths ($K$),
since multiple copies of the same species can react simultaneously.
However, this is not restrictive, since for $n=1$ the original definition of $T$ is recovered.
In addition to such stutter transitions, $T_{st}$ allows self loops at
terminal states, and therefore computations that require less than $K$ steps to reach a state satisfying $\phi_F$ can also be identified.

The encoding strategy described so far allows us to represent CRN synthesis as an SMT-problem and apply an SMT solver such as Z3 \cite{deMoura2008} to produce a CRN that satisfies the specification or prove that no such CRN exists for the choice of $M$, $N$ and $K$. More specifically, a solution CRN $\mathcal{C}$ is represented through the valuation of $\reac$ and $\prd$, which are extracted from the {\em model} returned by Z3.

In general, we are interested in enumerating many (or all possible) CRNs for the given class (defined by $M$, $N$ and $K$), which ensures that no valid solutions are omitted at that stage. To do so, we apply an incremental SMT-based procedure, where at each step we assert an uniqueness constraint guaranteeing that no previously discovered CRNs are generated. Given a concrete, previously generated CRN $\mathcal{C}' = (\Lambda, \mathcal{R}')$ and the new symbolic CRN $\mathcal{C} = (\Lambda, \mathcal{R})$ we are searching for (both of which are defined using the same species $\Lambda$), we define the constraint $\textit{DifferentFrom}(\mathcal{C}') \triangleq \neg \bigwedge_{r \in \mathcal{R}} \bigvee_{r' \in \mathcal{R}'} r = r'$, where $r = r'$ if and only if $\reac^r_s = \reac^{r'}_s \wedge \prd^r_s = \prd^{r'}_s$ for all $s \in \Lambda$. The new CRN $\mathcal{C}$ cannot simply be a permutation of the same reactions\footnote{At present, our uniqueness constraint does not consider other CRN isomorphisms but certain species symmetries are broken by the specification $\Phi_i$}. We start by generating a solution $\mathcal{C}'$ (if one exists), asserting the constraint $\textit{DifferentFrom}(\mathcal{C}')$, and repeating this procedure until the constraints become unsatisfiable, which corresponds to a proof that not additional CRNs exists for the given $N$, $M$, and~$K$.

\subsection{Tuning CRNs with parameter optimisation}
\label{sec:optim}

Here, we present our approach to optimising the reaction rates for CRNs satisfying $\{\Phi_0,\hdots,\Phi_n\}$. This becomes a parameter synthesis problem over a pCTMC set, analogous to parameter synthesis for a single pCTMC, as studied in \cite{Han2008,Ceska2014}. 
In contrast to this work, we aggregate over the multiple input combinations,
as specified in Problem~\ref{prob:main}.\ref{prob:secnd}.

To obtain solutions for the probability at a specified time $\pi_t$, we used numerical integration of the CME. Specifically, we used the Visual GEC software (\url{http://research.microsoft.com/gec}) to encode the CRNs and then integrate the CME for each combination of inputs.

To solve the maximisation problem, we used a Markov chain Monte Carlo (MCMC)
method, as implemented in the Filzbach software 
(\url{http://research.microsoft.com/filzbach}). 
Filzbach uses a variation
of the Metropolis-Hastings (MH) algorithm to perform Bayesian parameter
inference. The MH algorithm is used to approximate the posterior probability
of a parameter set from a hypothesised model taking on certain values,
constrained by a likelihood function. The probability of each parameter value
is then approximated by constructing a Markov chain of sampled parameter sets,
such that a proposed parameter set is accepted with some probability, based on
the ratio of the likelihood function evaluated at current and proposal
parameter sets. For more information on MCMC methods, see \cite{Robert2004}.
MCMC methods, such as simulated annealing, have also been shown to efficiently
find solutions to combinatorial optimisation problems \cite{Kirkpatrick1983},
taking a stochastic search approach similar to the MH algorithm. Stochastic
search can provide benefits over gradient-based optimisers by maintaining a
nonzero probability of making up-hill moves, protecting against getting stuck
in poor local optima. To use Filzbach for providing solutions to optimisation
CRN parameters, it is sufficient to encode the argument of
Problem~\ref{prob:main}.\ref{prob:secnd} as a
likelihood function. Subsequently, we generate MCMC chains
with suitably many burn-in iterations and samples to obtain an approximate
optimising parameter set~$\rates$.

\subsection{Calculating expected time}
\label{sec:time}

To evaluate the temporal performance of a CRN algorithm $\mathcal{C}$,
we make use of Markov chain theory to obtain the expected time until a terminal state is reached.
This is an exact measure of the expected running time for a given pCTMC with
inputs $i\in\mathcal{I}$, as opposed to using the mean of many stochastic
simulations~\cite{AAE2006}.

Let $A\subseteq X_r$ be the absorbing states of a pCTMC
$\mathcal{M}_p^\mathcal{C}=(X,\pi_0,\vec{Q}_p)$ and let $\tau^A$ be a vector
of expected hitting times, corresponding to the expected time of transitioning
from a state $x\in X_r$ to $A$. Then $\tau^A$ can be evaluated as the solution
to the equations~(page 113 of~\cite{Norris1997})
\begin{align*}
\tau_x^A = 0 & \text{ for } x \in A\\ %
- \sum_{x^\prime \in X_r} q_{x,x^\prime} \tau_{x^\prime}^A = 1 &\text{ for } x \notin A.
\end{align*}
Numerical solutions can be obtained by forming a matrix $W$ where the
rows and columns of $\vec{Q}_p$ corresponding to the terminal states ($A$) have been removed.
Then, $\tau^A$ is the solution to $W\tau^A=\vec{1}$, where $\vec{1}$ is the vector of 1's.
Numerical solutions can be obtained using Gaussian elimination.

Note that the time complexity analysis of CRNs typically assumes a volume $n$
equal to the maximum number of molecules in the system at any
time~\cite{CDS2012} (equivalent to parallel time in
PPs~\cite{AAE2006}).
This volume can be included by dividing each propensity by $n$ before
calculating expected time (see Section~\ref{sec:background}).
In the case of bimolecular CRNs this is equivalent to multiplying $\tau^A$ by $n$.

\section{Case studies}

  \subsection{Approximate majority}

\emph{Approximate Majority} is one of the most analysed functions in
distributed computing. It is the approximate version of the majority problem, which cannot be exactly computed by bimolecular CRNs (or population protocols) with less than 4 species~\cite{MNRS2014}.
For CRNs with 2 and 3 species there are known optimal (in terms of reaction
firings) approximate algorithms~\cite{Angluin2008,PVV2009}.

We specify the majority problem using the path predicate (see Section~\ref{sec:background}):
    $\Phi_{AM}(a,b) := \left(\phi_0(a,b), \phi_F(a,b)\right)$, where%
  \begin{align*}
    \phi_0(a,b) &:=
  \begin{cases}
  A = a \wedge B = b                & \mbox{ if } N=2, \\
  A = a \wedge B = b \wedge X = 0   & \mbox{ if } N = 3
  \end{cases} \\
    \phi_F(a,b) &:=
  \begin{cases}
      A^m_{a,b}  &\textrm{if } a > b\\
      B^m_{a,b}  &\textrm{if } a < b\\
      A^m_{a,b} \lor B^m_{a,b} & \textrm{otherwise}
  \end{cases} \\
    \text{ where } A^m_{a, b} &:= A =  a + b  \land B = 0 \text{ and }\\
     B^m_{a, b} &:=   A = 0  \land B =  a + b
  \end{align*}

\begin{figure}[ht]
\centering
\includegraphics[width=\linewidth]{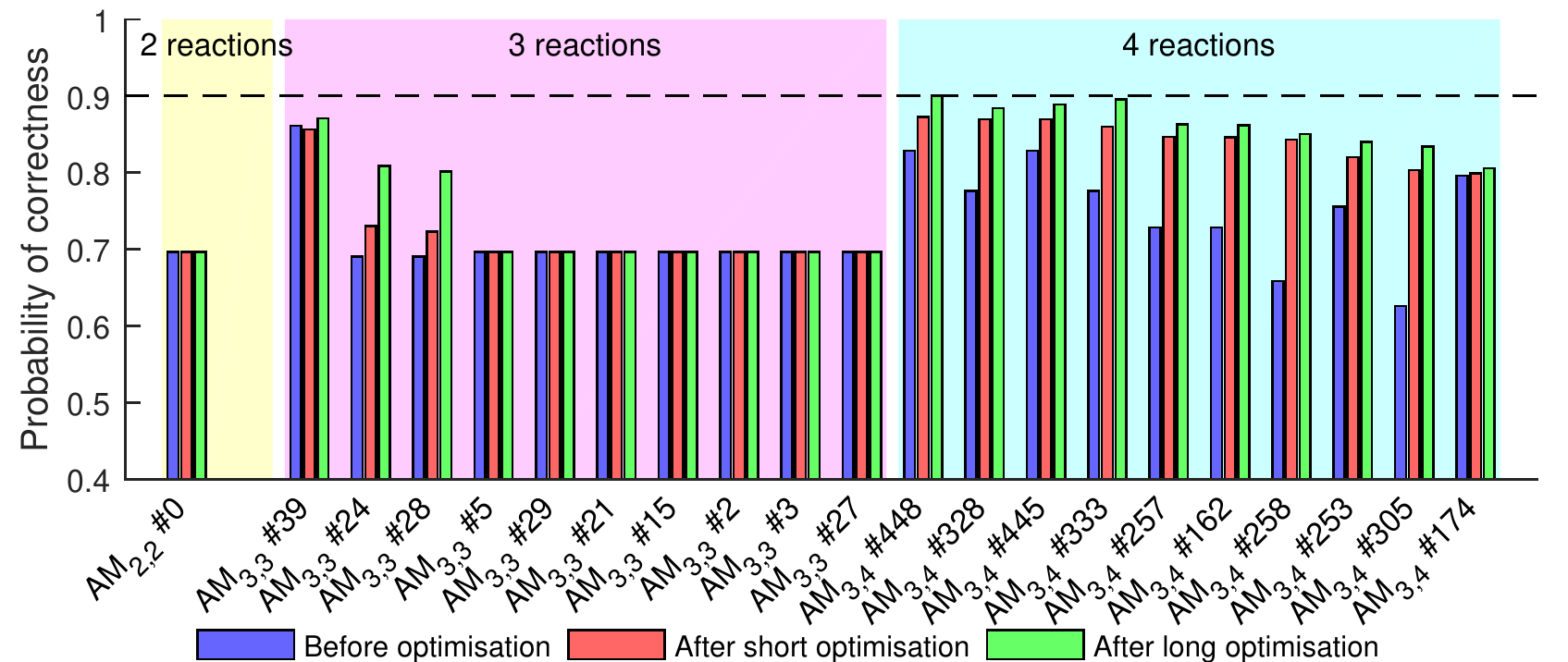}
\caption{\textbf{Performance of approximate majority circuits.} The SMT-based
  method was applied to the approximate majority specification for CRNs with
  2, 3 and 4 reactions. For each category, the top 10
  CRNs satisfying $\Phi_{AM}$ are ordered by their average probability
  after a short optimisation (20 burn-in, 20 samples; red bars).
  A longer optimisation (700 burn-in, 700 samples; green bars) was also
  performed. 
  We also show the average probabilities before optimisation (all rates equal
  to 1.0; blue bars).
  The dashed line is the average probability of CRN $AM_{3,4}$ \#448 after the
  longer optimisation, 0.8999, the maximum average probability in this trail.
}
\label{fig:AMhist}
\end{figure}

We used inputs $a,b \in [1\ldots 5]^2 \cup [6\ldots 10]^2$ for both optimisation and synthesis.
We applied the SMT approach to identify all CRNs with 2 to 4 reactions and 2
or 3 species that satisfy $\Phi_{AM}$ for $K \leq 5$ stutter steps (for $N$
species and $M$ reactions, there are $\binom{N^2 (N^2 - 1)}{M}$ total possible
CRNs). We used a short optimisation (20 burn-in, 20 samples) and sorted these
solutions by the value of $P_{\Phi_{AM}}$ for each. 
We then applied a longer optimisation (700 burn-in, 700 samples) to the
top 10 CRNs (Fig.~\ref{fig:AMhist}).

\begin{figure}[p]
\begin{minipage}[b]{0.2\textwidth}
\textbf{a} \quad AM$_{2,2}$ \#0
\scriptsize
\begin{align*}
A + B &\xrightarrow{82.8} B + B \\
A + B &\xrightarrow{82.9} A + A
\end{align*}
\vspace{1.5em}
\end{minipage}
\includegraphics[width=0.8\textwidth]{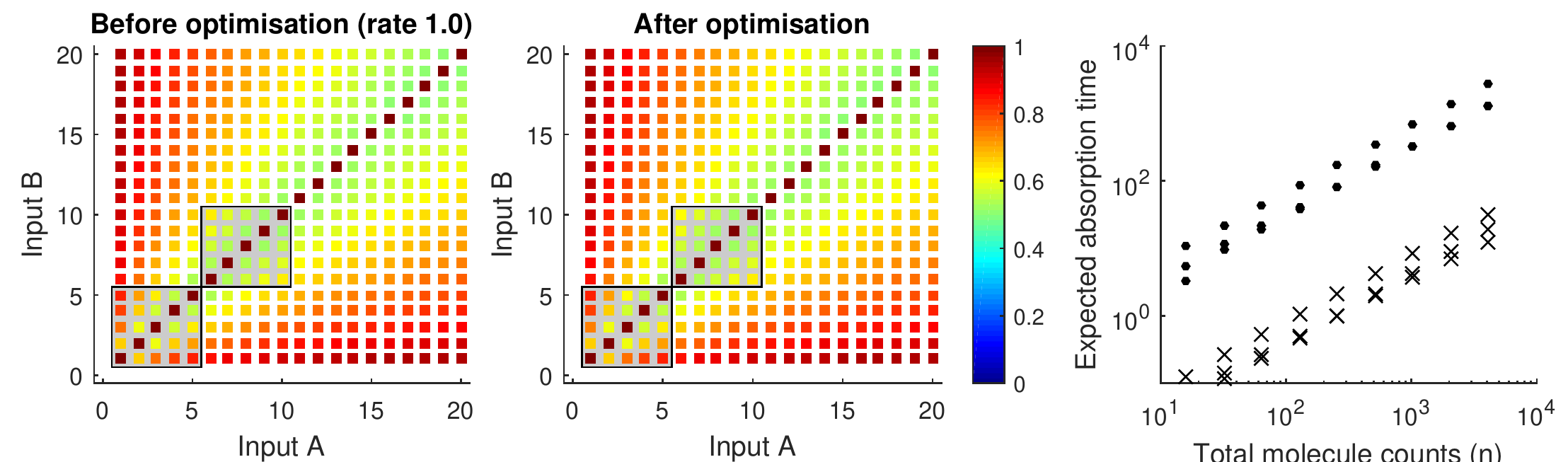}

\begin{minipage}[b]{0.2\textwidth}
\textbf{b} \quad AM$_{3,3}$ \#39
\scriptsize
\begin{align*}
A + B &\xrightarrow{92.9} X + X \\
A + X &\xrightarrow{26.2} A + A \\
B + X &\xrightarrow{23.3} B + B
\end{align*}
\vspace{0.5em}
\end{minipage}
\includegraphics[width=0.8\textwidth,trim=0 0 0 16,clip]{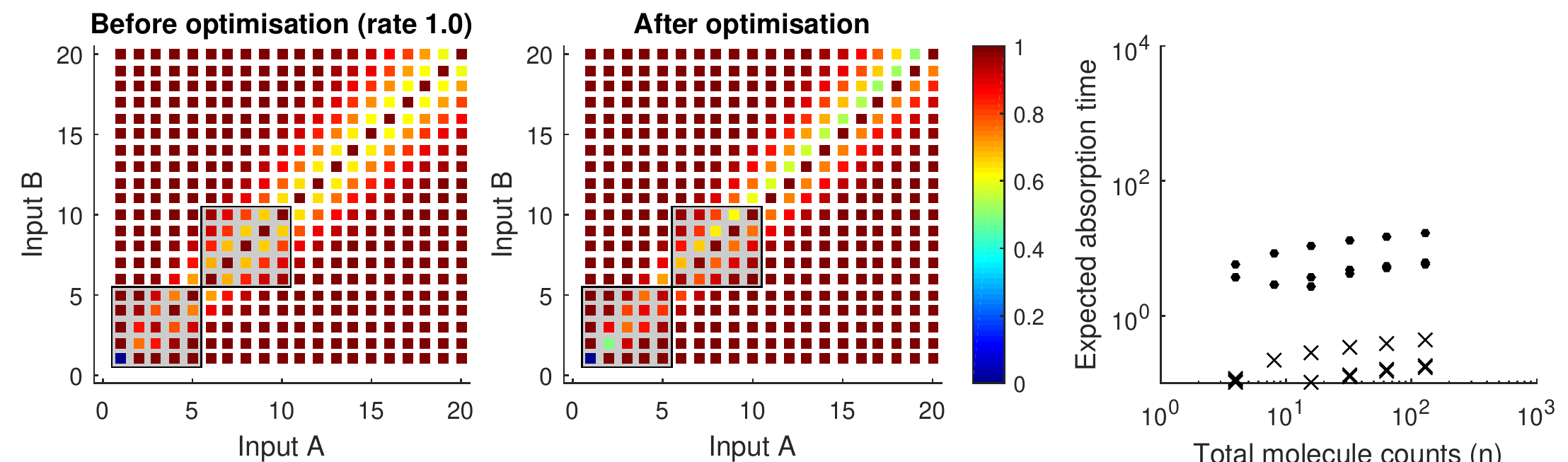}

\begin{minipage}[b]{0.2\textwidth}
\textbf{c} \quad AM$_{3,3}$ \#28
\scriptsize
\begin{align*}
B + X &\xrightarrow{6.40} B + B \\
X + X &\xrightarrow{0.89} A + A \\
A + B &\xrightarrow{35.9} X + X
\end{align*}
\vspace{0.5em}
\end{minipage}
\includegraphics[width=0.8\textwidth,trim=0 0 0 16,clip]{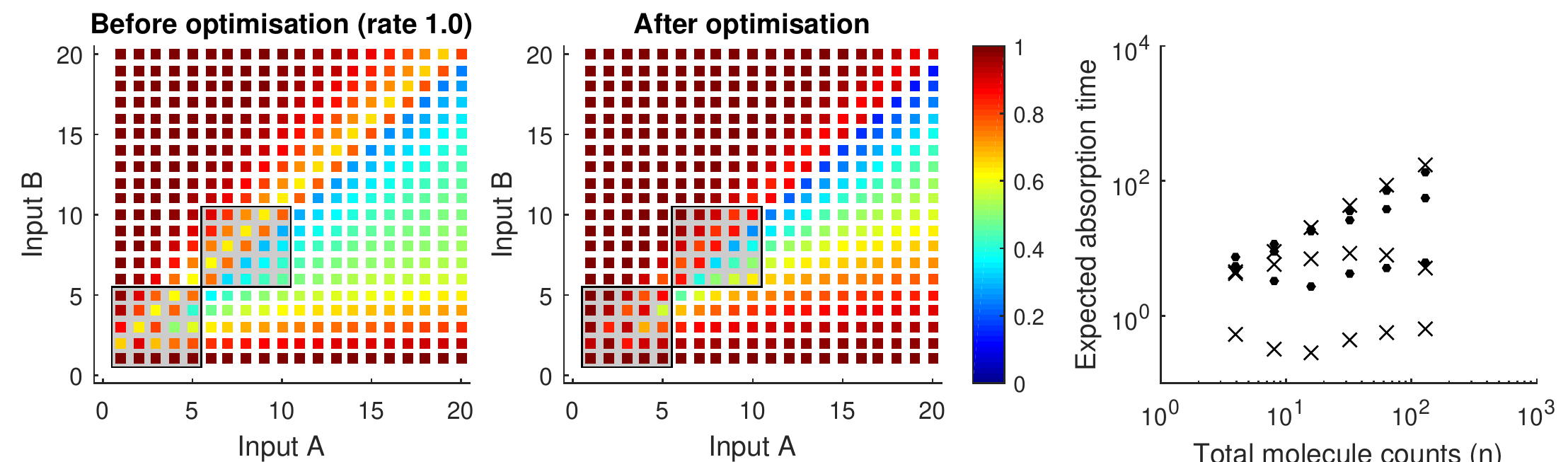}

\begin{minipage}[b]{0.2\textwidth}
\textbf{d} \quad AM$_{3,4}$ \#162
\scriptsize
\begin{align*}
B + X &\xrightarrow{17.9} B + B \\
A + B &\xrightarrow{65.6} X + X \\
A + B &\xrightarrow{0.34} A + X \\
A + X &\xrightarrow{16.5} A + A
\end{align*}
\vspace{0.5em}
\end{minipage}
\includegraphics[width=0.8\textwidth,trim=0 0 0 16,clip]{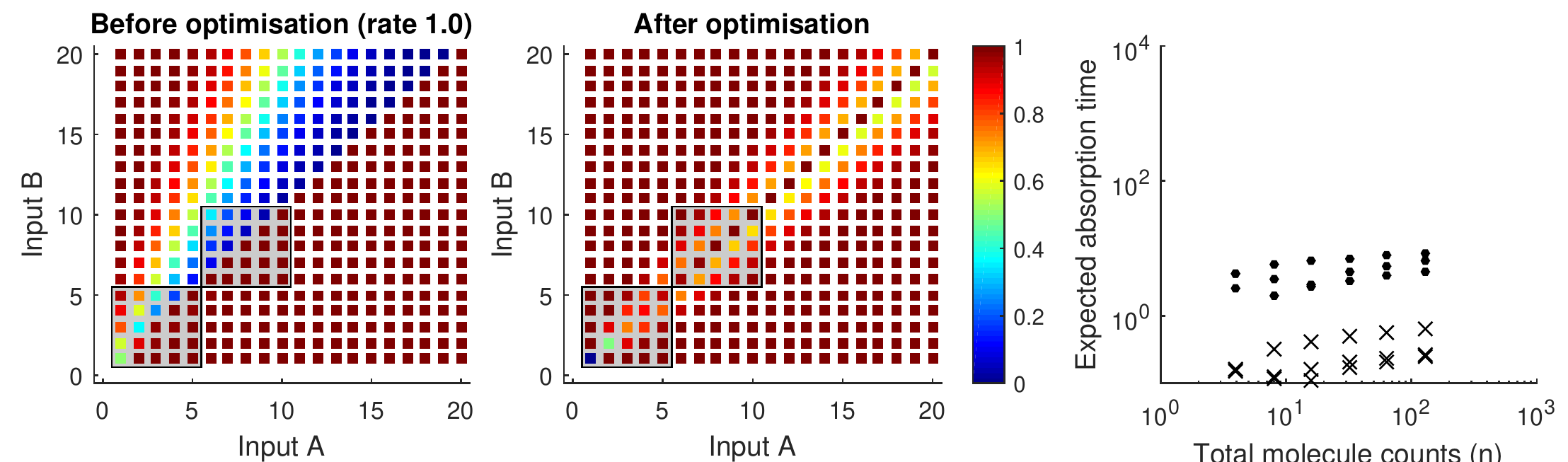}

\begin{minipage}[b]{0.2\textwidth}
\textbf{e} \quad AM$_{3,4}$ \#174
\scriptsize
\begin{align*}
B + X &\xrightarrow{80.6} B + B \\
A + B &\xrightarrow{11.6} B + X \\
A + B &\xrightarrow{56.6} A + X \\
A + X &\xrightarrow{3.91} A + A
\end{align*}
\end{minipage}
\includegraphics[width=0.8\textwidth,trim=0 0 0 16,clip]{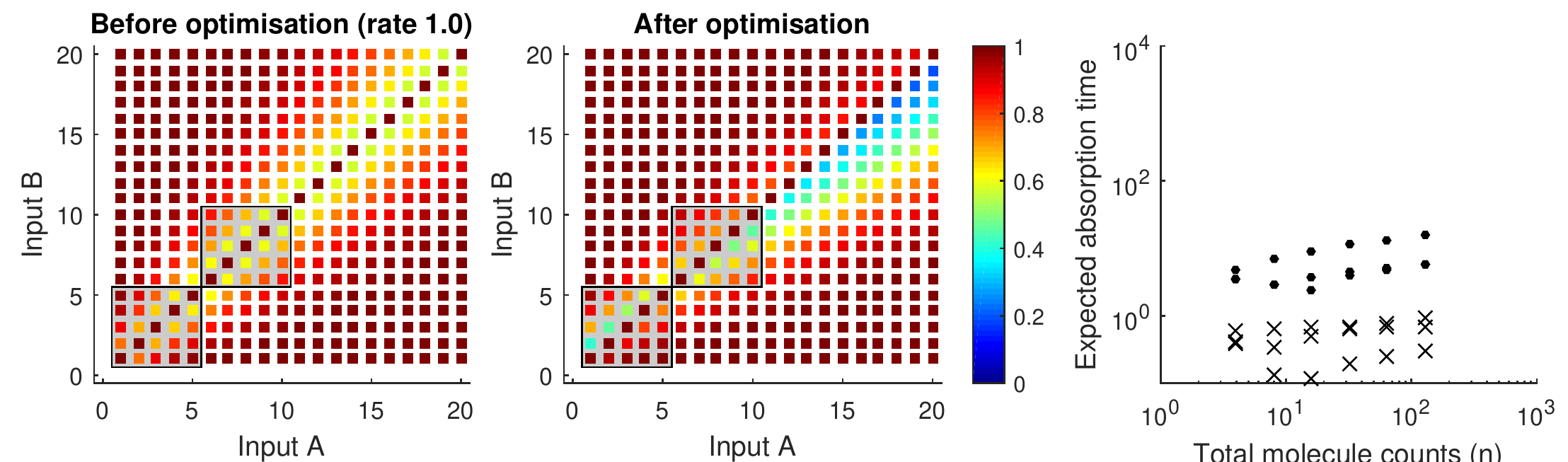}

\caption{\textbf{Response of Approximate Majority algorithms to varied
    inputs.} For each input combination, specified as initial copies of
  species $A$ and species $B$, the probability that both have the correct
  molecule count after 100 time units is reported. Results are shown for a
  variety of networks that performed well following optimisation (see \figref{fig:AMhist}). The performance of each CRN is compared both before
  optimisation (all rates equal to 1.0; left panels) and after long optimisation
  (central panels). 
  The grey boxes show the input ranges used for both generation and optimisation.
  The expected time until the CTMC reaches a terminal state
  is calculated for varying total molecule counts ($n$) (right panels).
  These times consider rates scaled as if occurring in a
  volume $n$ (see Section~\ref{sec:time}).
  The completion times for three alternative
  initial configurations (initial copies of $A$ were 10\%, 60\% and 90\% of $n$
  respectively) were calculated,
  illustrating minor differences in circuit completion times ($\times$ marks systems using optimised rates and \filledcirc marks systems using 1.0 for all rates).
}
\label{fig:AMheatmap}
\end{figure}

Using our approach, we found 1 CRN with 2 reactions and 2 species, the known \emph{direct competition} (DC) network~\cite{Cardelli2012} (\figref{fig:AMheatmap}a).
Out of 59,640 possible CRNs with 3 species and 3 reactions, the SMT solver
found 39 CRNs where $\Phi_{AM}$ was satisfied, 2 of which with probability
over 0.696 after the short optimisation (see \figref{fig:AMhist}).
These two networks ($AM_{3,3}$ \#24 and $AM_{3,3}$ \#28) are the dual of each
other and behave asymmetrically
but perform well owing to a compensatory asymmetric parameterisation (\figref{fig:AMheatmap}c).
One might expect that we should discover the known approximate majority
circuit~\cite{Angluin2008,Cardelli2014}, (see \figref{fig:AMheatmap}b).  %
However, this CRN does not satisfy the specification $\Phi_{AM}$ since,
for input $(A=1, B=1, X=0)$ the network terminates in the state $(A=0, B=0, X=2)$ and
thus fails to make a decision. If we remove this single problematic input from
the specification $\Phi_{AM}$, then this CRN is indeed discovered.
We include it for comparison as $AM_{3,3}$ \#39. 
Note that it scores a 0 on inputs $A=1$, $B=1$.

By increasing the number of reactions to 4, the SMT solver found
515 satisfying networks out of the 1,028,790 possible ones.
The top 5 networks, $AM_{3,4}$ \#448, \#328, \#445, \#333, and \#257 
have the same rules as the 3 reaction network $AM_{3,3}$ \#39 but each has 
a different 4th reaction. 
The network $AM_{3,4}$ \#162 had a lower performance than $AM_{3,3}$ \#39
before optimisation and was almost as good following optimisation.
This network was also asymmetric, with a corresponding asymmetric
parameterisation after optimisation (\figref{fig:AMheatmap}d).
The known 4 reaction network
$AM_{3,4}$ \#174~\cite{Cardelli2014} (\figref{fig:AMheatmap}e) is also
identified in 10th position.

Finally, we analysed the expected time until termination for each circuit,
using the procedure in Section \ref{sec:time} (right-hand panels of \figref{fig:AMheatmap}).
Note that Definition~\ref{def:prob} does not reward circuits 
that reach a high probability before the final time $t_f = 100$. 
However, in nearly all cases, the estimated hitting time of each system was improved by
optimisation.

\subsubsection{Computation times}
\begin{figure}[ht]
  \centering
  \includegraphics[width=\linewidth]{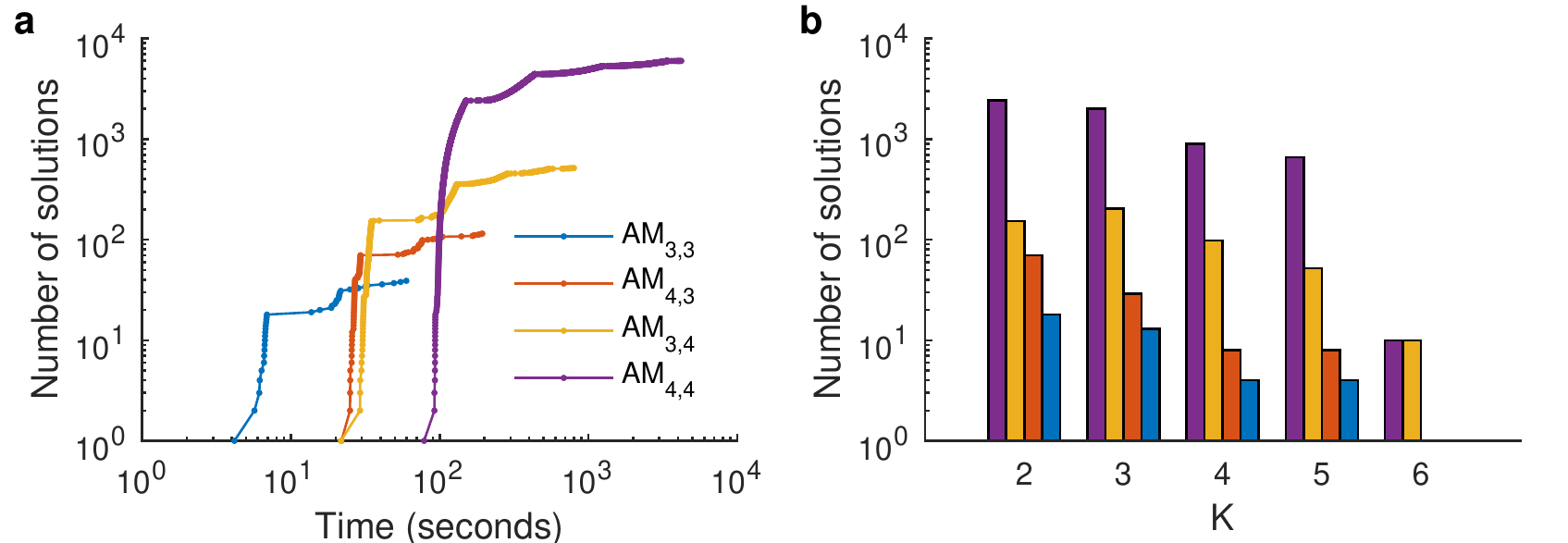}
  \caption{\textbf{Computation times for the SMT-based synthesis of majority decision-making CRNs}.
    Panel (a) shows the time required to generate a number of solutions (candidate CRNs)
    for $\Phi_{\mathrm{AM}}$ for $N$ species and $M$ reactions (denoted
    $AM_{N,M}$) for $N,M \in \{3,4\}^2$. 
    The computation was halted after 2 hours.
    Panel (b) shows the number of solutions found
    as $K$ (the length of considered computations with stutter transitions) increases.
  }\label{fig:Z3times}
\end{figure}

The computation times of our procedure depend on the size of the circuit ($M$ and $N$), length of considered computations ($K$) and exact specification $\Phi$ (including the number of given path predicates). We illustrate the computation times required for the SMT-based synthesis part
of our approach with the majority decision-making CRNs
(\figref{fig:Z3times}).

To determine how the CME calculation used in our method scales with
molecular copy numbers, we first ran calculations of the CME for the
established 3-reaction approximate majority CRN (system $AM_{3,3}$ \#39).
The calculation was initialised with $0.6n$ copies of $A$ and $0.4n$ copies of $B$, and all rates were set to 1.
As increasing the copy number decreases the simulation time interval over
which there are transient dynamics, we integrated the CME over the time
interval $\left[0, \frac{100}{n}\right]$, where $n$ is the total copy number.
We calculated transient probabilities at 500 output points, with $n\in[10, 1000]$.
This led to state-spaces of varying size, up to $10^6$, with all
calculations completing within 7200 seconds (2 hours). Smaller examples took
only a few seconds.
\begin{figure}[ht]
  \centering
\includegraphics[width=\linewidth]{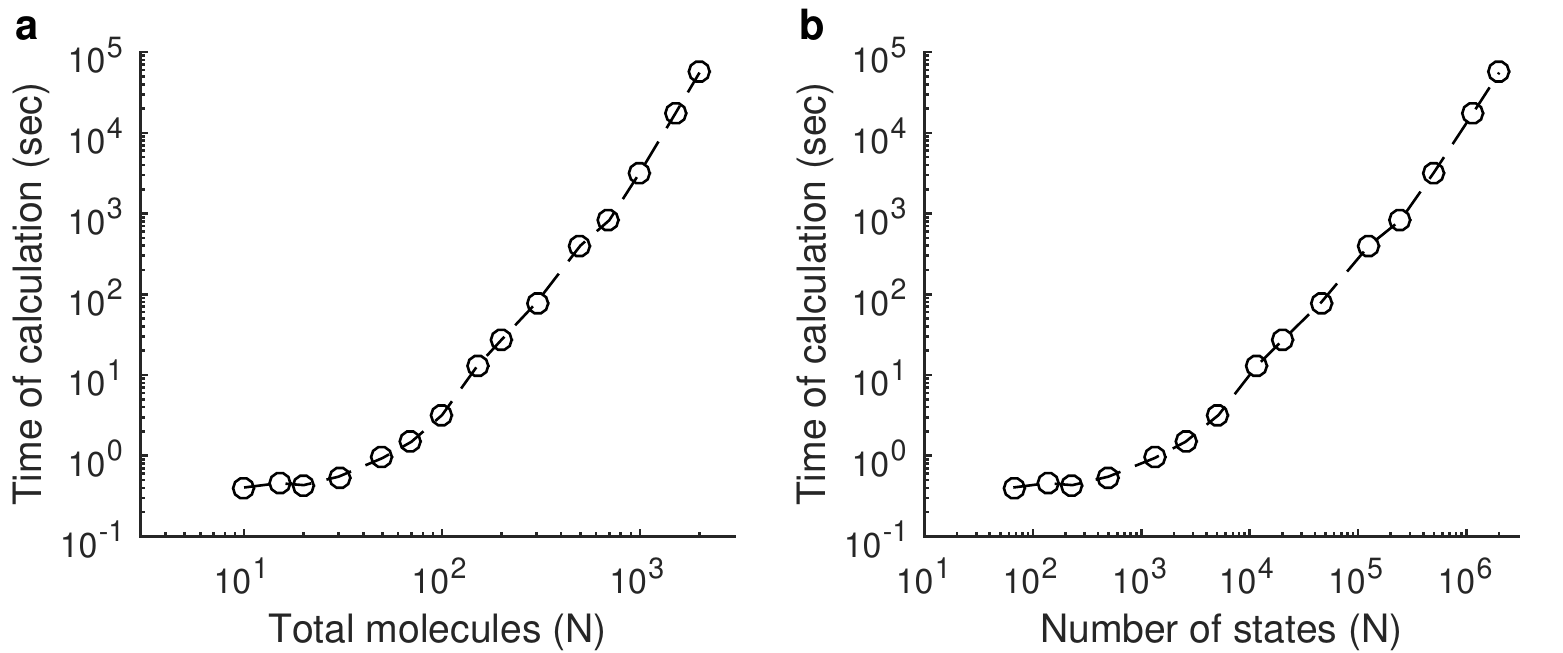}
\caption{\textbf{Transient probability calculation times for CRN} $AM_{3,3}$~\#39.
  Times indicated include the enumeration of the state-space,
  construction of a sparse matrix, then numerical integration in the interval
  $[0, \frac{100}{n}]$, where $n$ is the total molecule count. A single
  calculation was conducted for each value of $n$.}%
  \label{fig:CMEtimes}
\end{figure}

We can approximate the total run-time for parameter tuning as a function of the
number of iterations of the MCMC algorithm and the number of input
combinations assessed. For example, doing 200 iterations over 10 input
combinations which all have below 30 total molecules ($\lesssim$1 s each)
suggests a tuning procedure of no more than 2,000 seconds.

  \subsection{Division}
  
\emph{Division} is a non-semi-linear function and therefore it cannot be stably computed by CRNs~\cite{CDS2012}.
However, CRNs have been proposed that might implement the calculation of a ratio~\cite{Scialdone2013}, which allows plants to ration starch reserves during seasonally changing nights.

We specify the division problem using the path predicate
(see Section~\ref{sec:background}):
  \begin{align*}
    \Phi_{\mathrm{Div}}(a,b) &:= (\phi_0(a,b), \phi_F(a,b), \text{ where } \\
    \phi_0(a,b) &:=
      \begin{cases}
        A = a \land B = b \land X = 0 & \mbox{ if } N = 3\\
        A = a \land B = b \land X = 0 \land Y = 0 & \mbox{ if } N = 4
      \end{cases} \\
    \phi_F(a,b) &:= X = \left\lfloor \dfrac{a}{b} \right\rfloor
  \end{align*}
  
\begin{figure}[ht]
\label{fig:div_hist}
\centering
\includegraphics[width=0.95\linewidth]{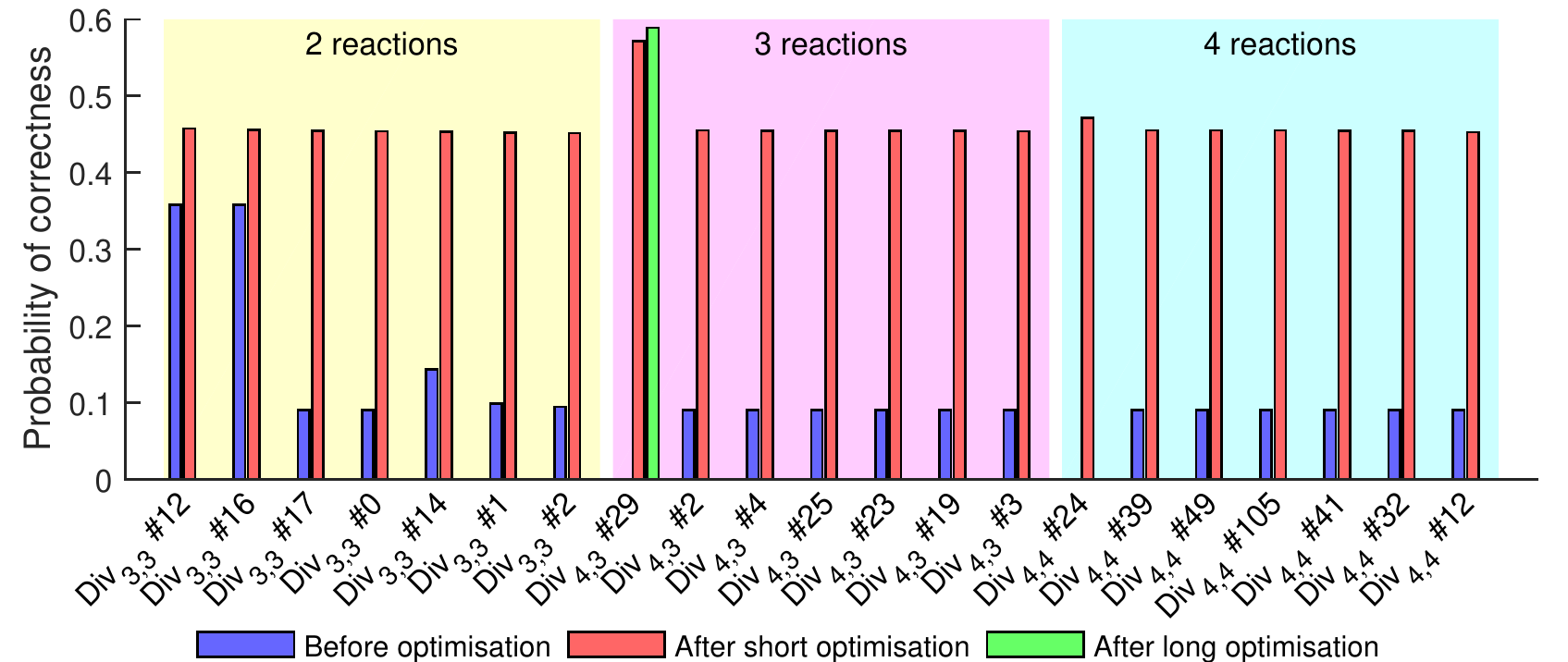}
\caption{\textbf{Performance of division circuits.}
  The SMT-based method was applied to the division specification for CRNs with 
  $N$ species and $M$ reactions for $N,M \in \{(3,3), (3,4), (4,4)\}$. 
  This figure shows the optimisation results for the top 7 CRNs in each
  category.
  The results are ranked and sorted by their average probability of being
  correct in the grey shaded zone after being optimised for 50 MCMC sample and
  burn-in steps (red bars).
  If a CRN scored an average probability of over 0.5 then it was optimised for
  a further 200 MCMC burn-in and sample steps.  
  The average probability is shown for satisfying CRNs before optimisation
  (all rates equal to 1.0; blue bars).}
\label{fig:Divhist}
\end{figure}

\begin{figure}[ht]
\begin{minipage}[b]{0.25\textwidth}
\textbf{a} \quad Div$_{3,3}$ \#12
\scriptsize
\begin{align*}
B + X &\xrightarrow{86.1} B + B \\ 
B + X &\xrightarrow{42.9} A + X \\
A + B &\xrightarrow{18.6} B + X 
\end{align*}
\vspace{1em}
\end{minipage}
\includegraphics[width=0.6\textwidth]{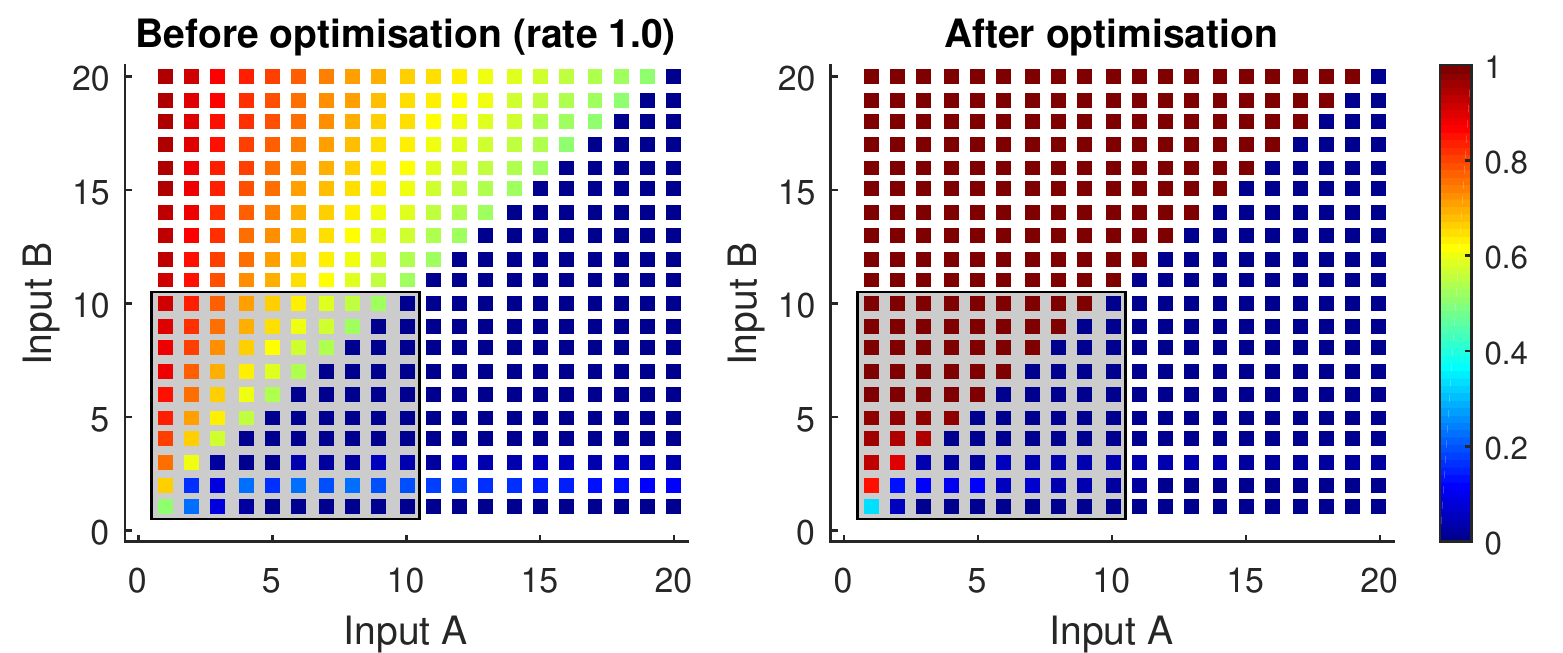}

\begin{minipage}[b]{0.25\textwidth}
\textbf{b} \quad Div$_{4,3}$ \#29
\scriptsize
\begin{align*}
B + X &\xrightarrow{87.5} B + Y \\
A + Y &\xrightarrow{0.15} X + Y \\
A + B &\xrightarrow{38.2} X + Y 
\end{align*}
\vspace{0.5em}
\end{minipage}
\includegraphics[width=0.6\textwidth,trim=0 0 0 16,clip]{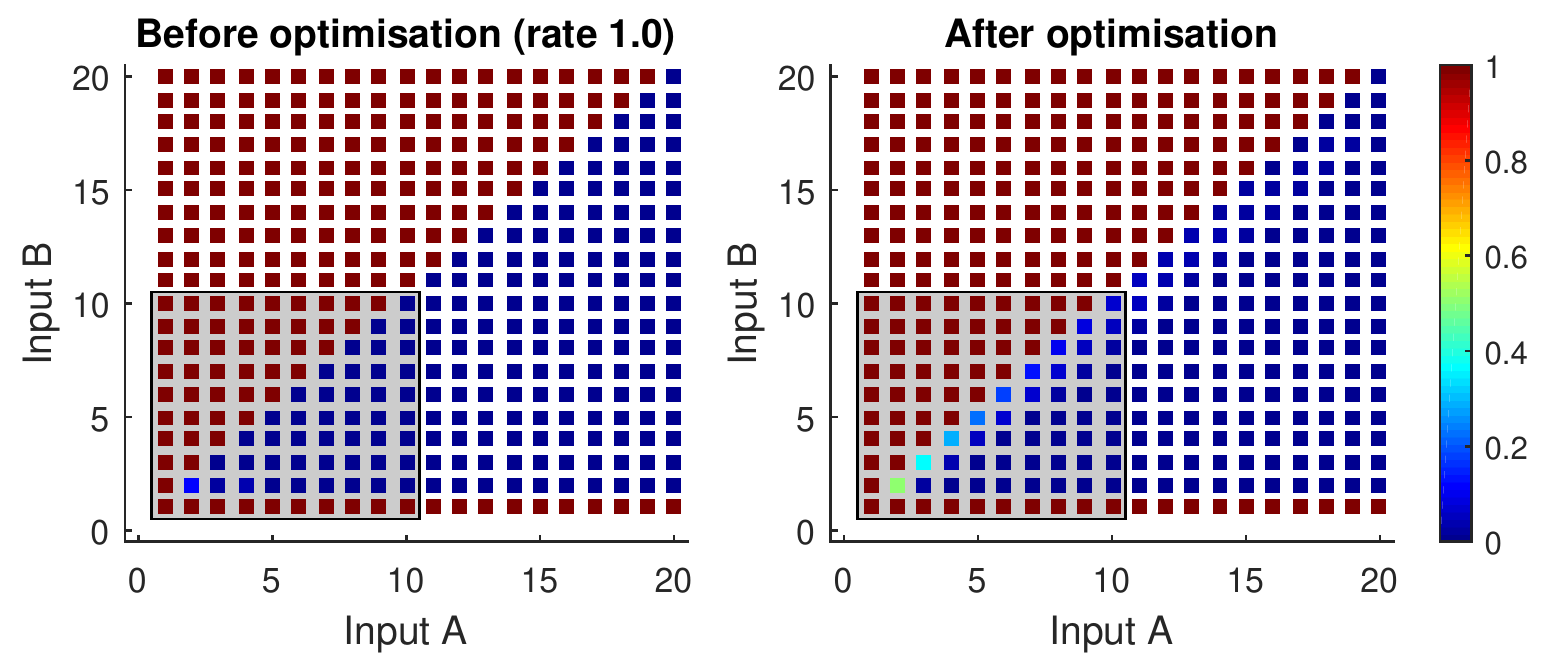}

\begin{minipage}[b]{0.25\textwidth}
\textbf{c} \quad Div$_{4,4}$ \#24
\scriptsize
\begin{align*}
A + Y &\xrightarrow{25.3} X + Y \\
A + X &\xrightarrow{34.9} X + X \\
A + B &\xrightarrow{1.3} X + Y\\
A + B &\xrightarrow{70.2} B + Y
\end{align*}
\vspace{0.5em}
\end{minipage}
\includegraphics[width=0.6\textwidth,trim=0 0 0 16,clip]{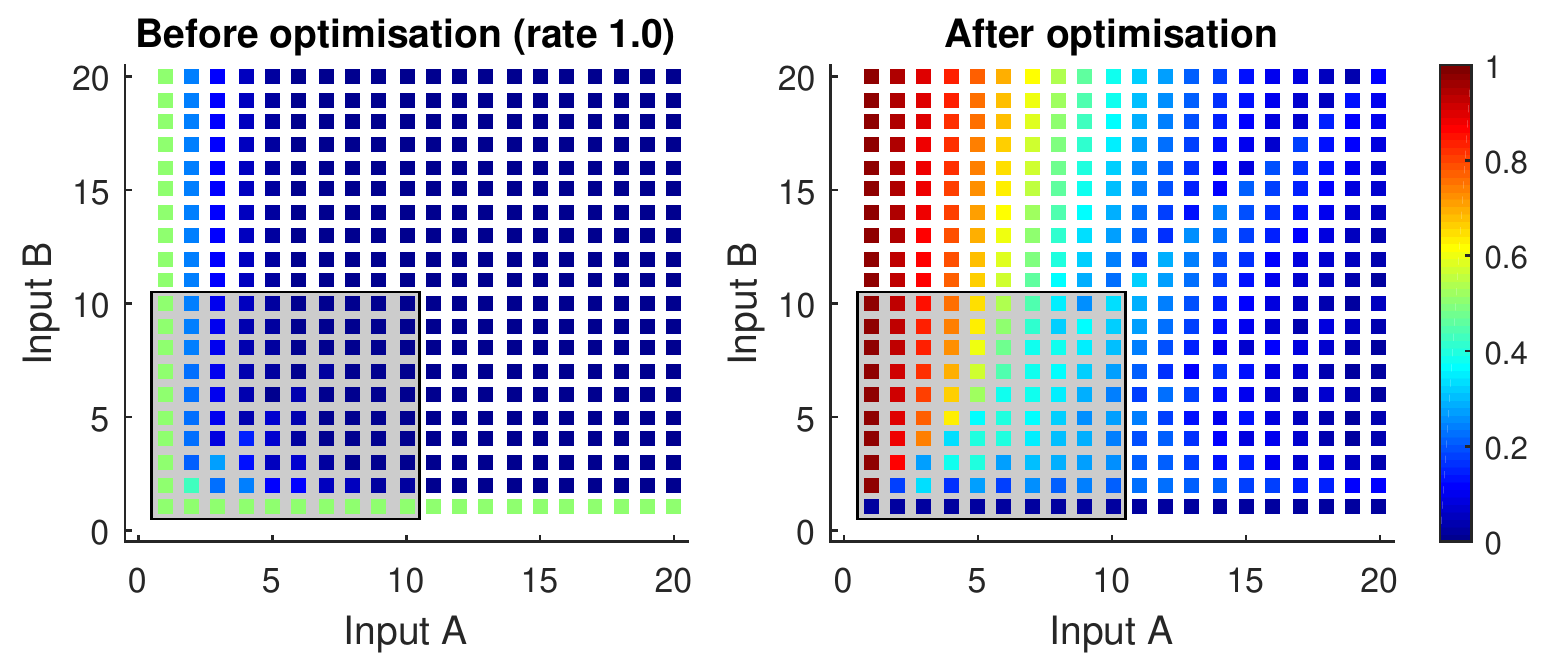}

\caption{\textbf{Response of Division algorithms to varied inputs.} 
  For each
  input combination, specified as initial copies of species $A$ and species
  $B$, the probability that the molecule count of $X$ is $\lfloor A/B\rfloor$ after 100 time units is reported.
  Results are shown for the top network in each combination of species
  and reactions (see \figref{fig:Divhist}).
  The performance of each CRN is compared both before optimisation (all rates equal to 1.0; left panels) and after optimisation (right panels).}
\label{fig:Divheatmap}
\end{figure}

We chose the input ranges $a,b \in [1,\ldots,10]^2$ for synthesis and optimisation
to give diverse selection of responses and to reinforce that 
$\lfloor\frac{a}{b}\rfloor=0$ when $a < b$. 
We applied the SMT approach to CRNs that satisfied $\Phi_{\mathrm{Div}}$ with
$K<20$ (without stutter transitions).
For 3 species and 3 reactions, 22 CRNs were discovered. 
For 4 species and 3 reactions, 34 CRNs were discovered. 
For 4 species and 4 reactions the first 105 CRNs were discovered. 
Of these, only one CRN $DIV_{4,3}$ \# 29 exceeded an average probability of 0.5,
though in most cases, optimisation improved performance substantially (\figref{fig:Divhist}).
For many of the generated circuits, high performance was observed only for $b > a$,
which should always evaluate to 0, with poor performance for the nonzero
output cases of $a > b$ (\figref{fig:Divheatmap}a,b).
Note that $Div_{4,3}$ \#29 is so far the top scoring divider CRN in this
class. 
Clearly, none of these circuits can be considered as \emph{good} algorithms for computing division, though our procedure was able to detect some very simple yet mediocre circuits in an automated way.
It is possible that better circuits will be found by considering CRNs with more reactions, species, and longer computation paths.

\section{Discussion}
In this paper, we presented a computational approach for the synthesis and parameter tuning of CRNs, given a specification  of the system's correctness.
We focused on the sub-class of bimolecular CRNs due to their importance as representations of various molecular algorithms and population protocols.
However, our approach is more general and could also be applied directly to
the synthesis of CRNs from other classes (e.g.\ unimolecular, trimolecular, etc.), which are defined through different stoichiometry constraints.
The CRNs we synthesize can be converted into equivalent physical
implementations, for example using DNA strand displacement
(DSD)~\cite{SSW2010,Chen2013}.  However, our approach could also be applied
directly to synthesize DSD systems through additional structural constraints.
This could lead to
simpler designs than the ones obtained through direct translation of CRNs.

We considered simple reachability properties defined in
 terms of predicates on the initial and final states of a computation which
 are sufficient to express various logical and arithmetic functions and operations.
 More general specifications, for example where intermediate states along computations are specified, are also currently possible within our approach but extensions to more expressive languages, such as the probabilistic temporal logics used with other methods~\cite{Ceska2014}, remains a direction for future work.

An alternative approach to the problem of realising arbitrary behaviour in
biochemical systems is to use directed evolution~\cite{Soyer2006,Dinh2014}
{\em In silico} evolutionary search strategies might scale to larger CRNs and
address the synthesis and parameter optimisation sub-problems using a single,
combined procedure. However, this comes at the cost of completeness, where the
absence of a solution does not mean a solution does not exist. In contrast,
our method addresses the sub-problems separately and uses the SMT solver and
theorem prover Z3 to identify CRNs that satisfy a given specification
(kinetics are ignored at this first stage). Since the results provided by Z3
are complete (for a sufficiently large $K$), the termination of the procedure
with no solutions is a ``proof'' that no CRNs exist in the given class.
Thus, besides providing a practical tool for the identification of CRNs with
given behaviour, the completeness property
means our approach could also help explore the
theoretical limits of CRN computation (e.g. no CRNs with less than $M$
species and $N$ reactions that compute a given function exists).
For many applications, elements of our method could be
complementary with evolutionary algorithms.
For example, the exact CTMC methods we use to assess the
probability of correct computations in a given CRN could provide a useful
fitness function for evolutionary search, compared to alternative approximate
methods based on stochastic simulation.

The fully automated generation of ``good'' CRNs is a challenging problem and certain scalability limitations of our current method must be addressed to provide a more complete solution.
Firstly, the SMT-based synthesis procedure we propose may represent large or infinite state spaces and handle systems with large molecule numbers.
However, currently this method is limited to relatively small CRNs with few reactions, species, and which have short computation paths.
Secondly, the CTMC methods we apply require an explicit representation of the state space, which must be finite (which is always the case for biomolecular CRNs initialised with a finite number of molecules) and contain few reachable states --- this makes the method suitable for systems involving relatively few species and numbers of molecules.
To circumvent the need for an explicit representation of the state space, stochastic dynamical behaviour could be approximated by averaging multiple trajectories from Gillespie's stochastic simulation algorithm \cite{Gillespie1977}, using fluid or central limit approximations \cite{Ethier2009}, or using ordinary differential equations. Depending on the specification, and the nature of the CRN, some of these approaches might be appropriate, but none are free of their own documented limitations.
Finally, the large number of solutions identified at the synthesis stage of
our approach makes the parameter tuning phase challenging and indicates that
additional constraints describing more accurately the structure and dynamics
of ``good'' solutions could improve the method.

For tuning reaction rates, alternative cost functions could be
  used that reward solutions that are ``nearly" correct, e.g. using a
  mean-squared error. This would be most appropriate in high copy number
  situations, where a precise number of molecules is not integral. Our
  approach is more appropriate for systems operating at low copy numbers,
  offering an exact characterisation of the probability that a specific
  predicate is satisfied. Our results were shown for calculations at $t_f=100$
  time units, a transient probability, rather than at the stationary
  distribution. While the selection of $t_f$ is subjective, it allows a
  circuit programmer to specify how long they are willing to wait for a
  computation. Circuits that reach high probability at $t>t_f$ will not be
  rewarded. However, a natural extension to the presented method would be to
  reward circuits that reach high probability at $t<t_f$, both imposing an
  upper bound on time and optimising within that range. This could be achieved
  by integrating our metric over the interval $[0, t_f]$.

Automating the search for CRNs that compute the solution for a specified
problem would be beneficial to both theoretical and experimental molecular
programmers.
Our method can be used to show the existence or absence of CRNs of a certain
size and also suggest CRNs that %
can be tuned for a specific input range, and so become candidate designs for experimental construction.
Prior to construction, more in-depth analysis of the candidate CRNs produced is beneficial, including parameter sensitivity/robustness analysis and bifurcation analysis (where appropriate). Future work could also incorporate notions of robustness into the proposed method, for example by using interval-based methods \cite{Ceska2014}.
Our results illustrate the potential of this approach on several examples,
including the majority and division functions discussed here.

\subsection*{Acknowledgements}
We thank Dan Alistarh and Luca Cardelli for helpful discussions on the development and applications of our methodology.

\bibliographystyle{ieeetr}
\bibliography{biblio}

\clearpage

\end{document}